\documentclass[prb,twocolumn]{revtex4-1}

\usepackage{amsmath}
\usepackage{graphicx}
\usepackage{setspace}
\pdfoutput=1
\begin{document}

\title{Numerical modeling of nonohmic percolation conduction and Poole Frenkel laws }
\author{Maria Patmiou}
\email{maria.patmiou@rockets.utoledo.edu} 
\affiliation{Department of Physics and Astronomy, University of Toledo, Toledo, OH 43606, USA}
\author{V. G. Karpov}
\email{victor.karpov@utoledo.edu}
\affiliation{Department of Physics and Astronomy, University of Toledo, Toledo, OH 43606, USA}
\author{G. Serpen}
\email{gursel.serpen@utoledo.edu}
\affiliation{Electrical Engineering and Computer Science, University of Toledo, OH 43606, USA}
\author{B. R. Weborg}
\email{Brooke.Weborg@rockets.utoledo.edu}
\affiliation{Electrical Engineering and Computer Science, University of Toledo, OH 43606, USA}

\date{\today}

\begin{abstract}
We present a numerical model that simulates the current-voltage (I-V) characteristics of materials that exhibit percolation conduction. The model consists of a two dimensional grid of exponentially different resistors in the presence of an external electric field. We obtained exponentially non-ohmic I-V characteristics validating earlier analytical predictions and consistent with multiple experimental observations of the Poole-Frenkel laws in non-crystalline materials. The exponents are linear in voltage for samples smaller than the correlation length of percolation cluster $L$, and square root in voltage for samples larger than $L$.

\end{abstract}

\maketitle

\section{Introduction}

\begin{figure}[t!]
\includegraphics[width=0.32\textwidth]{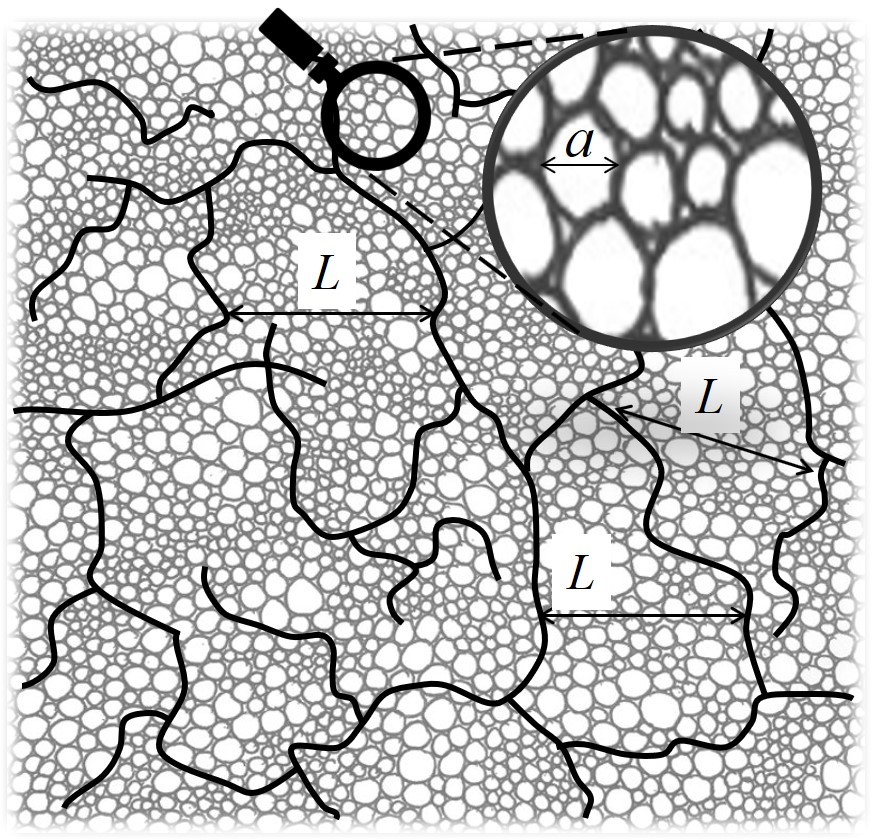}
\caption{A fragment of the infinite cluster (representative of e. g. a polycrystalline structure) with correlation radius $L\gg a$. \label{Fig:IC}}\end{figure}

Non-ohmic conduction in disordered materials can be described in terms of the percolation  theory. \cite{efros,shklovskii1979,levin1984,patmiou2019} Materials that exhibit percolation conduction include amorphous, polycrystalline and doped semiconductors, and granular metals. According to the percolation theory, the electric current flows along an infinite conducting cluster with topology resembling that of waterways formed in a flooded mountainous terrain.\cite{efros1986,stauffer1994} Each cluster bond consists of $L/a\gg 1$ exponentially different microscopic non-ohmic resistors in series, where $L$ is the correlation radius of the cluster (its mesh size) and $a$ is the linear dimension of one resistor as illustrated in Fig. \ref{Fig:IC}. The material is effectively uniform for length scales above $L$.

An intuitively transparent case of percolation conduction is presented by a system of thermally activated resistors with resistances $R_i=R_0\exp(-W_i/kT)$. The  activation barriers $W_i$ are random, and the exponents $\xi _i=W_i/kT$ vary in a broad interval $\Delta \xi\approx \xi _m \gg 1$, in which their probabilistic distribution $\rho (\xi )$ is approximately uniform. \cite{efros,shklovskii1979,levin1984,patmiou2019} Here $\xi _m$ is the upper boundary of the distribution. Since the number of markedly different activation energies is of the order of $\xi _m$ and each cell of the cluster must include all representative resistors, one can estimate
\begin{equation}\label{eq:L}L\sim a\xi _m.\end{equation}

\begin{figure}[bht]
\includegraphics[width=0.47\textwidth]{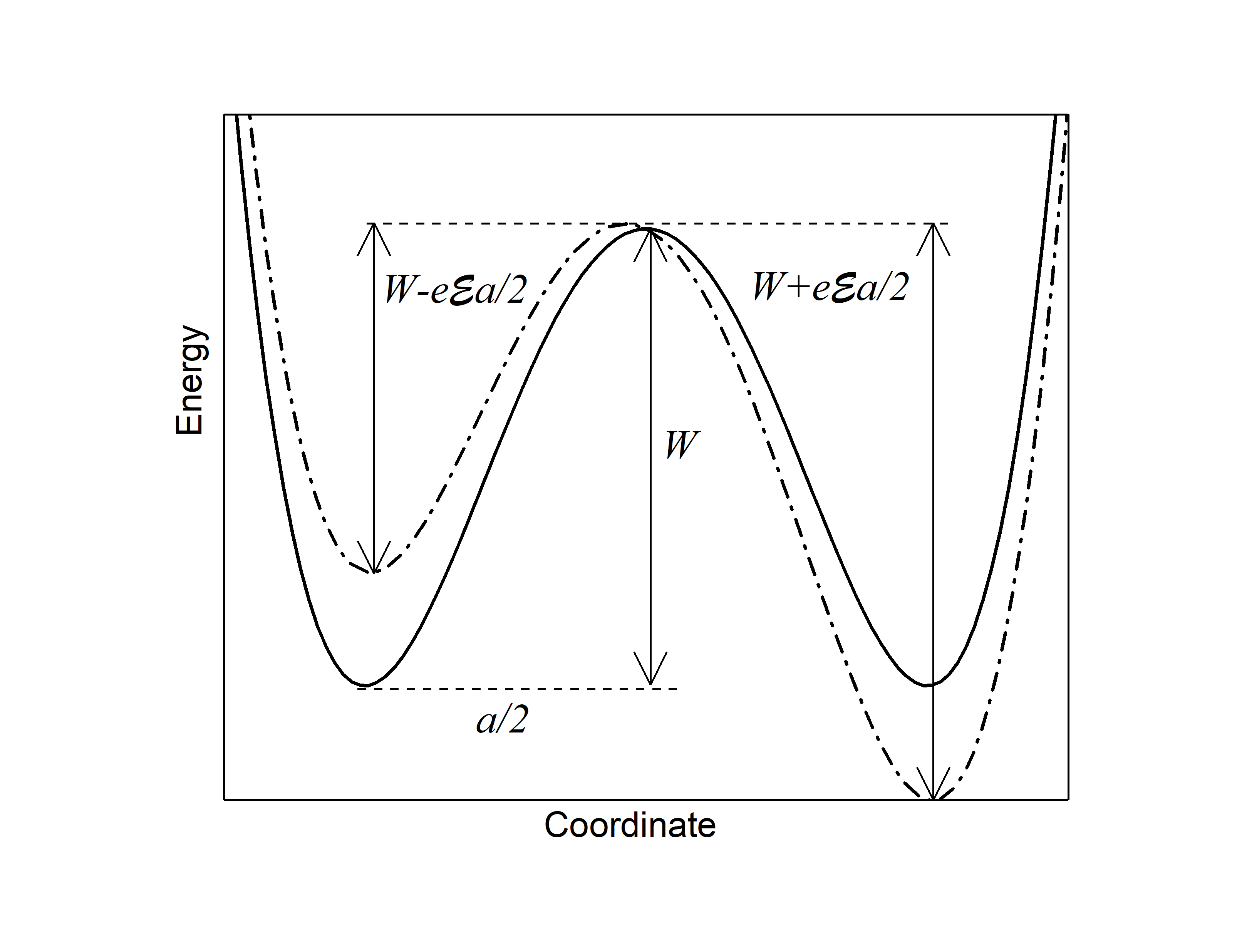}
\caption{Sketch of a two-center energy diagram illustrating the energy barrier and its change under external field ${\cal E}$. \label{Fig:DWP}}\end{figure}

It is customary to describe the non-ohmic resistors by their field dependent currents $J_i({\cal E}_i)$ in the form,
\begin{equation}\label{eq:nonohm}J_i\propto\exp\left(-\frac{W_i}{kT}\right)\sinh\left(\frac{ae{\cal E}_i}{2kT}\right)\end{equation} where $k$ is Boltzmann's constant,  $T$ is the temperature, $e$ is the electron charge, and ${\cal E}$ is the field strength. The term $\sinh(ae{\cal E}_i/2kT)$ in Eq. (\ref{eq:nonohm}) represents the sum of currents $\exp(\pm ae{\cal E}_i/2kT)$ along and against the field as illustrated in Fig. \ref{Fig:DWP}. The  activation barriers $W_i$ are random, and their corresponding exponents $\xi _i=W_i/kT$ vary between different resistors in a broad interval $\Delta \xi\gg 1$, in which their probabilistic distribution $\rho (\xi )$ is approximately uniform. \cite{efros,shklovskii1979,levin1984,patmiou2019}

Because the current in a cluster's bond must be the same for all its constituting resistors, i. e. $-W_i+ae{\cal E}_i/2 = const$, the voltages ${\cal E}a$ across these resistors are distributed non-uniformly, concentrating on the strongest (highest  $W_i$) resistors, which enhances the non-ohmicity. The percolation theory predicts the macroscopic conductivity of the form
\begin{equation}\label{eq:FP}\sigma ({\cal E})=J/{\cal E}\propto \exp[f({\cal E})],\end{equation} with
\begin{equation}\label{eq:F}f({\cal E})=C\sqrt{{\cal E}}/(kT) \quad {\rm and }\quad C=\sqrt{cqaW_0}\end{equation}
where $W_0$ is the characteristic energy scale of the random potential, $a$ represents its length scale attributable to the linear dimension of one resistor, and $c\sim 1$.

An attempt of numerical modeling \cite{levin1984} did not confirm Eq. (\ref{eq:F}), which failure could be attributed to that era's (1984) computer resources insufficient for modeling a large system of non-linear random resistors. In a recent paper, \cite{patmiou2019} Eq. (\ref{eq:F}) was derived using an approach different from that of the original work. \cite{shklovskii1979} In addition, it was stated there that for the case of `small' samples with linear dimensions below $L$, Eq. (\ref{eq:F}) should be replaced with
\begin{equation}\label{eq:P}f({\cal E})= C'{\cal E}/(kT) \quad {\rm and}\quad C'=aq/2. \end{equation}

Apart from the above outlined theoretical problematics of transport in disordered systems, Eqs. (\ref{eq:F}) and (\ref{eq:P}) (with different coefficients $C$ and $C'$) remained for about a century on the forefront of condensed matter research as the renowned Frenkel-Poole laws describing electric conduction in a broad variety of materials; here, we point at two reviews \cite{nardone2012,schroeder2015} and a few recent of more than hundred of other publications. \cite{ielmini2007,ismail2015,kim2014,hu2014,huang2014,schulman2015,slesazeck2015,song2018,lim2000,yuan2017}

The original derivations of Frenkel-Poole (FP) laws \cite{poole1916,hill1971,frenkel1938,perel} assumed the mechanism of field induced decrease of the ionization energy of a single Coulomb trap (Frenkel) or a pair of such (Poole), which is rarely applicable in the range of actual temperatures and fields. \cite{nardone2012,schroeder2015,perel} Most importantly though remains the fact that FP laws are not observed in materials where they should be most applicable, doped crystalline semiconductors, yet they show up consistently in noncrystalline materials where the Coulomb centers, if present, should form disordered arrays.

Based on the latter observation, it is logical to assume that the disorder and its related percolation conduction form the natural base for the understanding of FP. Furthermore, validation of the analytically derived  Eqs. (\ref{eq:F}) and (\ref{eq:P}) for disordered systems becomes increasingly important. With that goal in mind, here, we present the numerical modeling of non-ohmic conduction in a system of random resistors described by the current-voltage characteristics of Eq. (\ref{eq:nonohm}).

\section{Modeling}\label{model}

\begin{figure}[b!]
\includegraphics[width=0.5\textwidth]{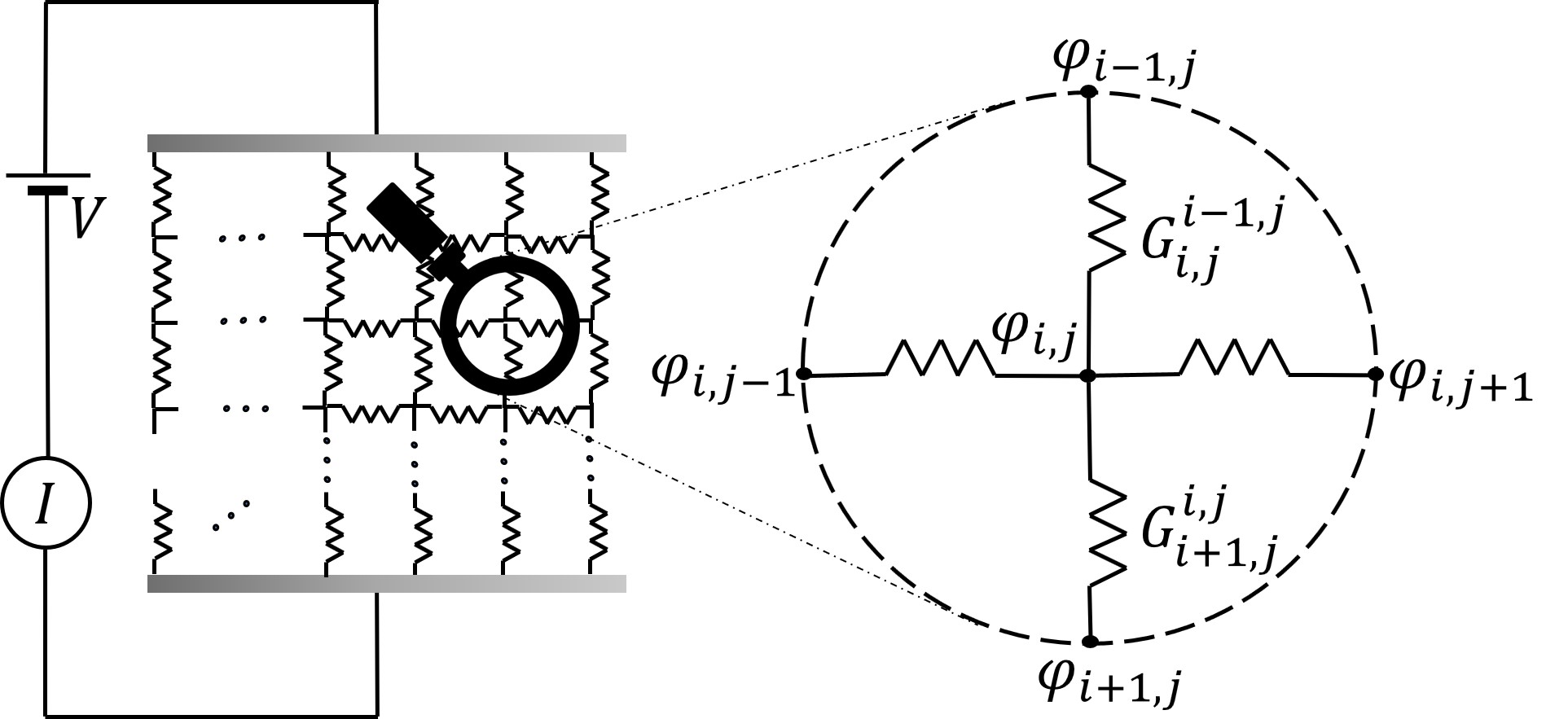}
\caption{NxN grid of random resistors, where N is the number of nodes on each side. The nodes on the side of the grid depicted on the top of the figure are at voltage $V$, while the opposite electrode is taken as the reference electrode. \label{Fig:resgrid}}\end{figure}

We are modeling a square 2-D grid of non-ohmic random resistors as shown in Fig. \ref{Fig:resgrid}. Our algorithm described next is the same as developed by Levin \cite{levin1984} in his 1984 publication which  showed only a linear log-current versus voltage dependence. We are taking advantage of better computing capabilities that allow us to expand the size of the grid and the dispersion of the resistances beyond Levin's work. A formal complication with notations arises because of the 2-D geometry where each node is described by two indexes ($x$ and $y$-coordinates), so that the inter-node quantities require 4 indexes. For example,  Eq. (\ref{eq:nonohm}) can be utilized to represent the current between two neighbor nodes (i,j) and (i,j+1) where the voltage drop between the nodes is:
\begin{equation}U_{i,j+1}^{i,j}= \varphi_{i,j} -\varphi_{i,j+1} +{\cal E}a.\end{equation}
Here and below, the upper (superscript) and lower (subscript) indexes describe the transitions respectively from and to the nodes.

Along the same lines, we introduce the following notations:
\begin{equation}f_{i,j}=\exp\left(\frac{e\varphi_{i,j}}{2kT}\right),  \quad S_{i,j+1}^{i,j} =\exp\left(\frac{e{\cal E}a}{2kT}\right),\nonumber \end{equation}
\begin{equation}S_{i,j-1}^{i,j}=\exp\left(\frac{-e{\cal E}a}{2kT}\right),  \quad S_{i+1,j}^{i,j}=S_{i-1,j}^{i,j}=1,\nonumber \end{equation}
\begin{equation}G_{i,j+1}^{i,j}=\exp\left(-\xi_{i,j+1}^{i,j}\right), \quad \xi_{i,j+1}^{i,j}\equiv W_{i,j+1}^{i,j}/kT.\nonumber \end{equation}

where $W_{i,j+1}^{i,j}$ is the barrier for a transition from the node $(i,j)$ to the node $(i,j+1)$.

Using the above notations, Eq. (\ref{eq:nonohm}) can be presented in the form,
\begin{equation}\label{eq:curden}J_{i,j+1}^{i,j}=
\frac{G_{i,j+1}^{i,j}}{2}\Big( \frac{f_{i,j}S_{i,j+1}^{i,j}}{f_{i,j+1}}-  \frac{f_{i,j+1}}{S_{i,j+1}^{i,j}f_{i,j}}\Big)\end{equation}
We note that the conductances are symmetric: $G_{k,l}^{i,j}=G_{i,j}^{k,l}$ since from the microscopic point of view they are related to the height of a single potential barrier separating the equivalent (in zero field) potential barrier as illustrated in Fig. \ref{Fig:DWP}. The electric field is a function of the external voltage: ${\cal E}=(V/Na)$, and room temperature is assumed, $kT/e\sim0.025V$.

From Kirchhoff's first law, the algebraic sum of the currents on each node equals zero. Using Eq. (\ref{eq:curden}) and the above given substitutions, we obtain the following expression:
\begin{equation}\label{eq:nodalv}f_{i,j}=\Bigg[\frac{\left(\displaystyle\sum_{k,l} G_{k,l}^{i,j} f_{k,l}/S_{k,l}^{i,j} \right)}{ \left(\displaystyle\sum_{k} G_{k,l}^{i,j}S_{k,l}^{i,j}/f_{k,l} \right)}\Bigg]^{1/2}\end{equation}
where the summation over $k,l$ involves the 4 neighbors nearest to $(i,j)$ node (except for $i=1$ and $i=N$, where only 3 neighbors have to be considered.

The random system of resistors was introduced by generating random quantities $\xi_{i,j+1}^{i,j}$ in the interval $(0,\xi _m)$ where $\xi_m\gg 1$. Our initial approximation was that $\varphi_{i,j}=0$ except $\varphi_{i,j}$ related to the nodes on the left and right electrodes that remain at voltages $V$ and 0 respectively. Eq. (\ref{eq:nodalv}) was used to iterate the quantities $f_{i,j}$.

  At the end of each iteration, the total current $I(V)$ was calculated as a sum of all currents determined by Eq. (\ref{eq:curden}) through a cross-section perpendicular to the direction of the electric field. Our initial convergence criterion which required that two successive iterations led to a percent error less than 0.2 percent \cite{levin1984} was not sufficient to guarantee the electric charge conservation. We therefore implemented the convergence criterion
\begin{equation}\label{eq:Delta}
\Delta =\left[\frac{1}{N^2}\sum _{i,j}\frac{\left(\sum _{k,l}J_{k,l}^{i,j}\right)}{\sum _{k,l}\left(J_{k,l}^{i,j}\right)^2}\right]^{1/2}
\end{equation}
where, similar to Eq. (\ref{eq:nodalv}), the summation over $k,l$ involves the 4 neighbors nearest to $(i,j)$ node. The quantity $\Delta$ represents the relative rms of the currents and the criterion of convergence guaranteeing that Eq. (\ref{eq:nodalv}) is solved with sufficient accuracy $\delta$ requires $\Delta \lesssim \delta$.
Some details of our programming are explained in the Appendix in terms of the principal pseudocode. The pseudocode in its entirety can be found in the Supplementary material while the pseudocode conventions are covered in reference [26].

\section{Modeling results}\label{sec:res}

Our modeling was performed concurrently with MATLAB (R2019b), Python (3.7) and C++(Visual C++ 2017)
platforms, all leading to the same results. Data utilized in the figures were gathered with MATLAB (unless otherwise indicated). We obtained a series of current-voltage characteristics for NxN 2-D grids with a range of $N$ between 5 and 200 and the disorder parameter $\xi _m$ in the range between $\xi _m=1$ to $\xi _m=20$. The convergence times for the code were increasing exponentially as $N$ and/or $\xi _m$ increased as shown in tables \ref{tab:conv_times} and \ref{tab:conv_times_C} .

\begin{figure}[h!]
\includegraphics[width=0.47\textwidth]{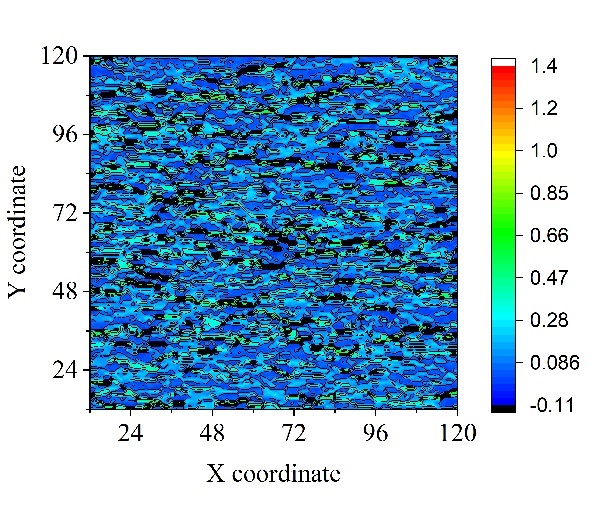}
\caption{Percolation current pathways in a large (120x120) grid for $\xi _m=10$ and $V=10$ V forming a random mesh topology. Rare black domains represent reverse currents. The currents are presented in relative units \label{Fig:map2}}\end{figure}

\begin{table}[h!]
\caption{Examples of computational convergence times -using MATLAB- for various grid sizes and different values of the parameter $\xi _m$.}
\begin{tabular}{|c|c|c|}
\hline
N (nodes) & t (s) for $\xi _m =10$ & t(s) for $\xi _m =20$\\
\hline
10& 0.9-2.5 \footnotemark[1] &0.3-3.9 \\
40& 25-29  & 1725-2171\\
70& 194-277 &failed \footnotemark[2] \\
100& 184-192 &failed \\
\hline
\end{tabular}
\footnotetext[1]{The given ranges represent convergence times for three different seeds. The computations were made using an Intel(R) Core(TM) i7-8700 CPU @ 3.2 GHz,12.0 GB with Windows 10 operating system.}\footnotetext[2]{The designation failed refers to a lack of obtaining the first value of the current after a period of 30 minutes.}
\label{tab:conv_times}
\end{table}
 \begin{table}[h!]
\caption{Examples of computational convergence times -using C++- for various grid sizes and different values of the parameter $\xi _m$.}
\begin{tabular}{|c|c|c|}
\hline
N (nodes) & t (s) for $\xi _m =10$ & t(s) for $\xi _m =20$\\
\hline
10& $\sim$0.05 \footnotemark[1] &0.03-0.08 \\
40& 1.9-2.1  & 46.9-111.8\\
70& 11.0-11.5 &741.9-1183.7  \\
100& 12.-18.9 &3527.8-6328.2 \\
200&248.3-308.7&failed\footnotemark[2]\\
\hline
\end{tabular}
\footnotetext[1]{The given ranges represent convergence times for the same three seed values as those in table (\ref{tab:conv_times}). The computations were made using an Intel(R) Core(TM) i7-7820HQ CPU @2.90GHz, 16 GB Ram, with Windows 10 Enterprise operating system.}\footnotetext[2]{Test not run due to large quadratic growth in runtime for $\xi_m=20$.}
\label{tab:conv_times_C}
\end{table}
For definiteness we present our findings here for the case of $\xi _m=10$. According to Eq. (\ref{eq:L}) a cell of the infinite percolation cluster will encompass roughly a 10x10 nodes fragment of our structure in Fig. \ref{Fig:resgrid}. We then discriminate between small $N\times N$ systems with $N<10$ and large $N\times N$ systems with $N\gg 10$. More specifically we chose $N=5$ and $N\sim 100$ to represent our findings about the current flow and statistics in small and large systems respectively.  We observed that:

\begin{enumerate}
\item For grids that range approximately between $N=10$ and $N=30$ the dependence of $\ln I$ vs. $V$ is best fit by a linear function as shown in Fig. \ref{Fig:lin_fit} for the case of $\xi_m=10$.
\begin{figure}[h!]
\includegraphics[width=0.47\textwidth]{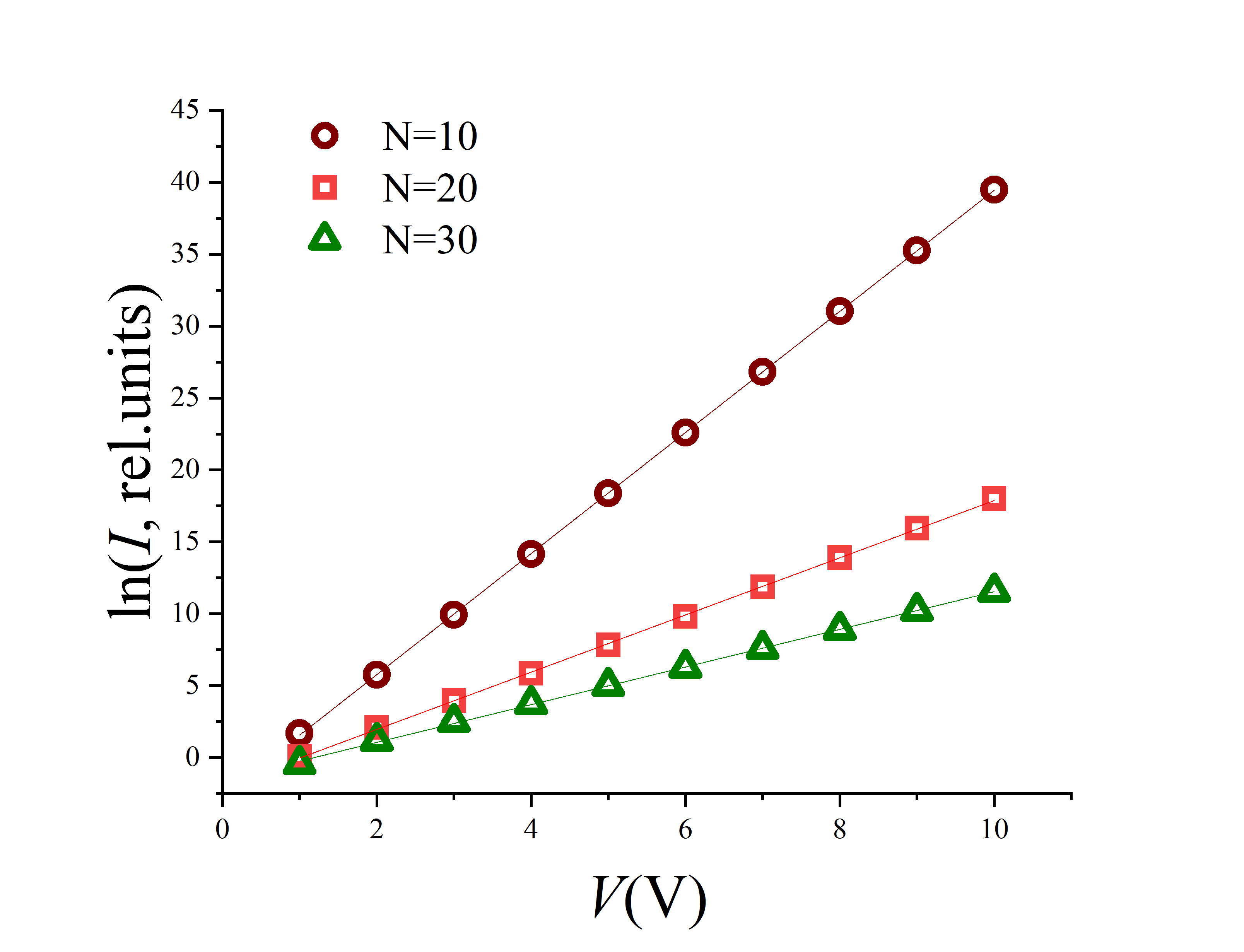}\caption{Current-Voltage characteristics for small grids with $\xi_m=10$. The solid lines represent a straight line fit.\label{Fig:lin_fit}}\end{figure}
\item Higher orders of $N$ produced a `square root' dependence $\ln I = const + \sqrt{V}$ illustrated in Fig. \ref{Fig:sqrt_fit}.
\begin{figure}[h!]
\includegraphics[width=0.47\textwidth]{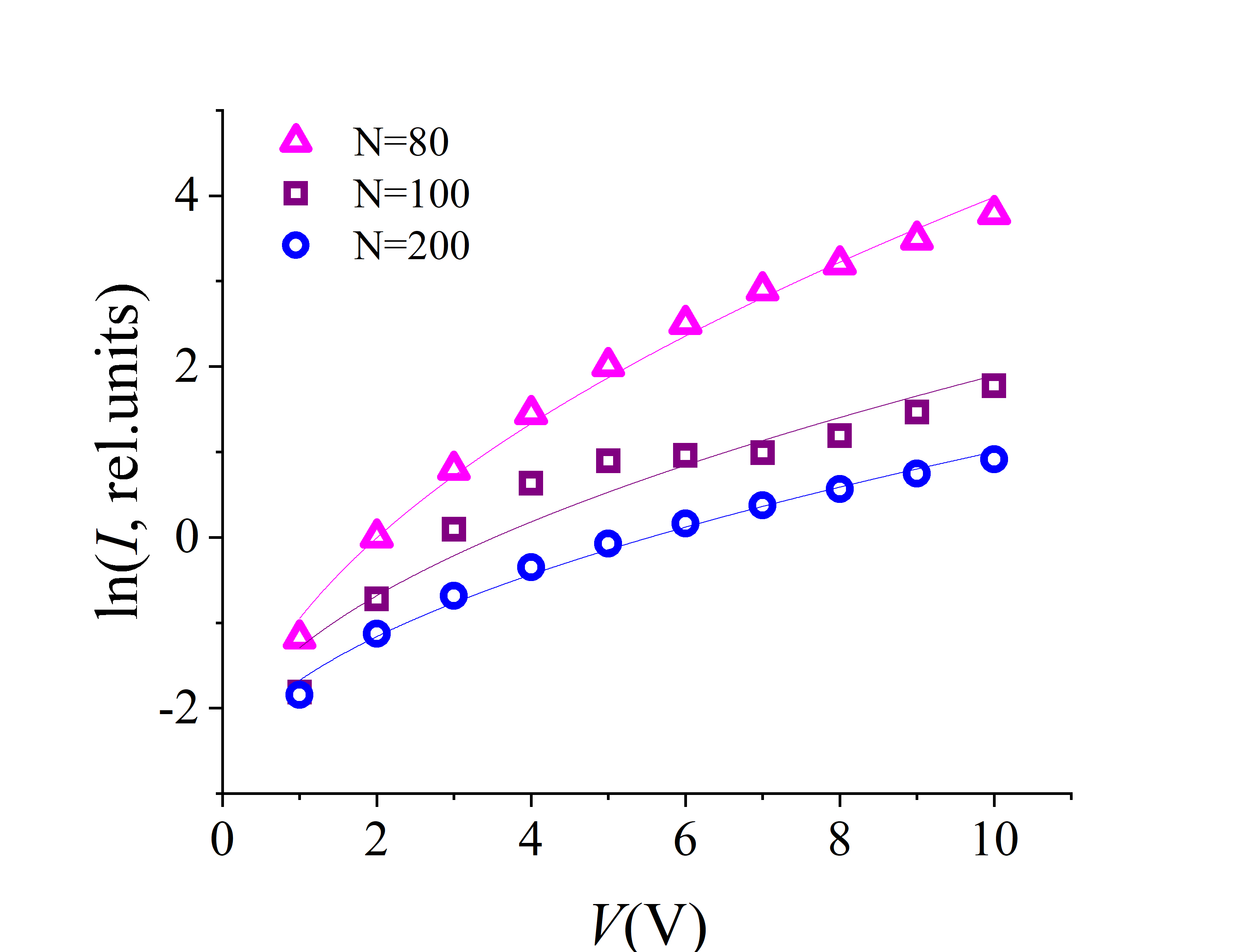}\caption{Current-Voltage characteristics for a system with $\xi_m=10$. The solid lines represent a square root fit.(Data for this figure were obtained using C++.) \label{Fig:sqrt_fit}}\end{figure}
\item In the intermediate range of $N\approx 40$  to $N\approx 70$ both the quality of either linear or square root fits are less perfect than in the examples of Figs. \ref{Fig:lin_fit} and \ref{Fig:sqrt_fit}.  In general, the transition between the linear and the square root dependence is not sharp, and occurs over a range of sizes.
\item There are statistical outliers in the above trends (i.e. the $N=100$ characteristic shown in Fig. \ref{Fig:sqrt_fit}). The general trends remain the same for  a range  of $\xi_m$ between 1-10.
\item The percolation conduction pathways appear to be close to rectilinear (`pinholes') in small grids with linear dimensions below the correlation length $L$. Their particular configurations vary between nominally identical random samples as illustrated in Fig. \ref{Fig:map1}.
    \begin{figure}[t!]
\includegraphics[width=0.25\textwidth]{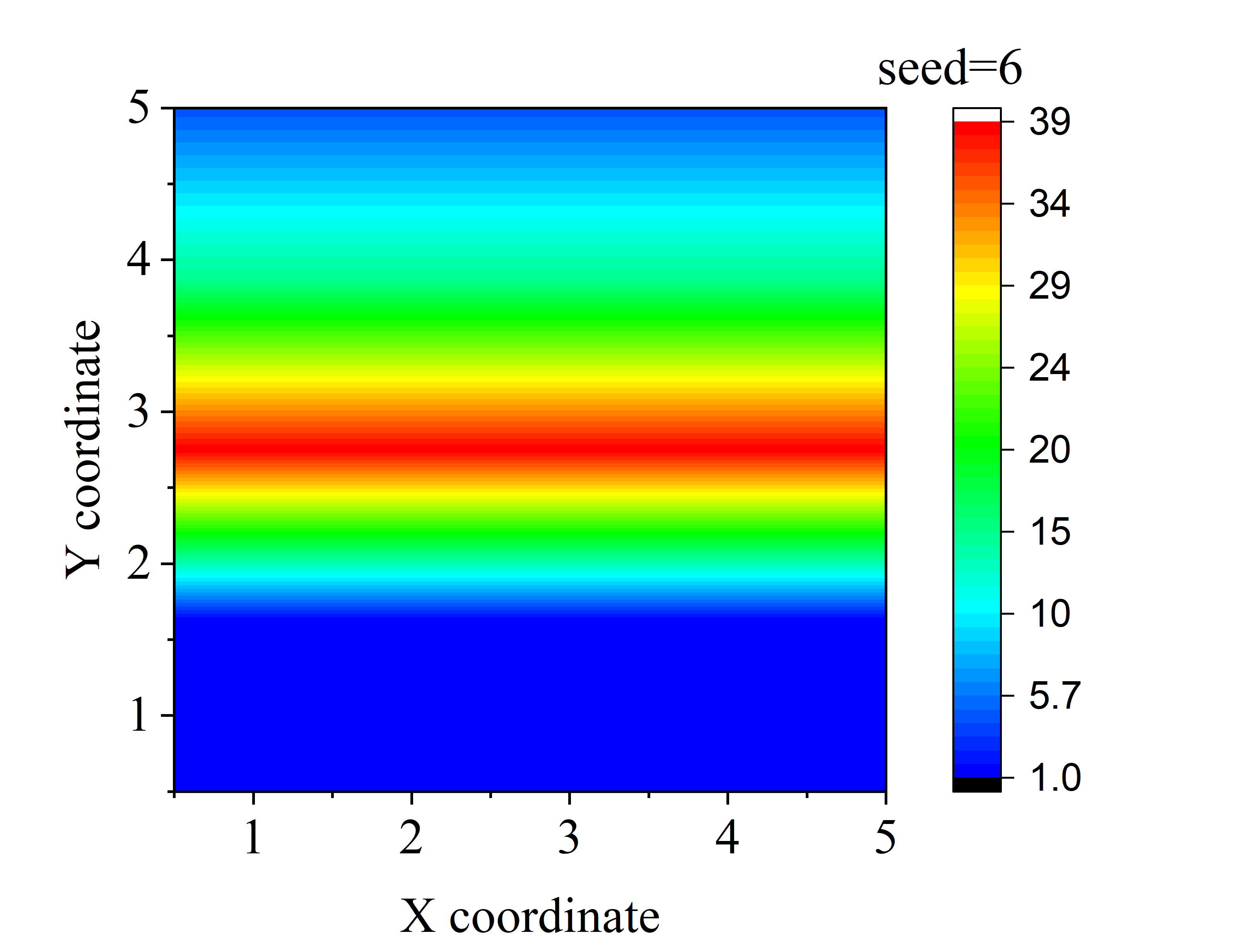}\includegraphics[width=0.25\textwidth]{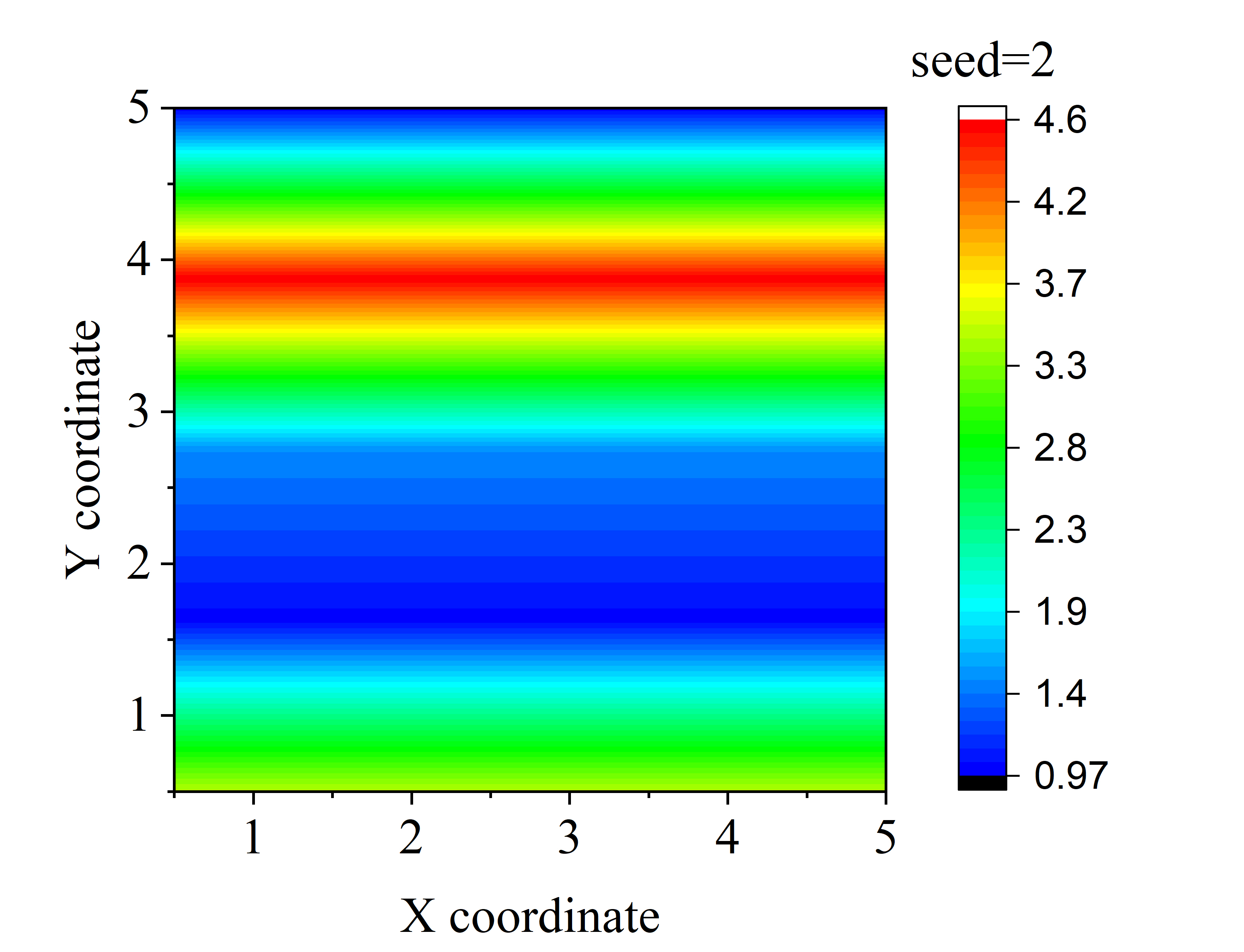}
\caption{Electric current distributions in two nominally identical small (5x5) grids with different disorder configurations (seeds) between two vertical electrodes. The currents are presented in relative units. \label{Fig:map1}}\end{figure}
\item The statistics of currents in small grids can be satisfactory fit by the log-normal distributions as illustrated in Fig. \ref{Fig:lognormals}.
    \begin{figure}[h]
\includegraphics[width=0.25\textwidth]{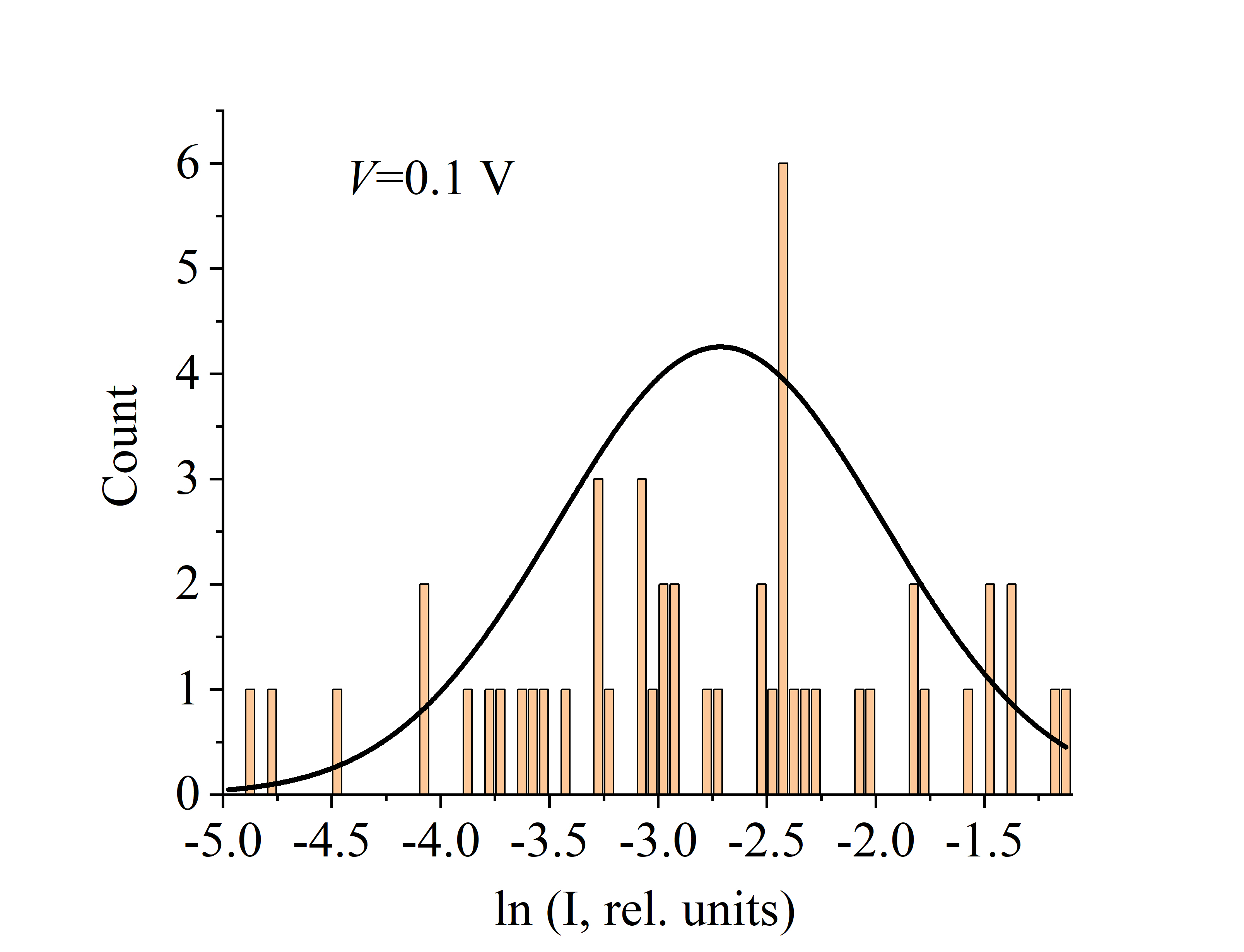}\includegraphics[width=0.25\textwidth]{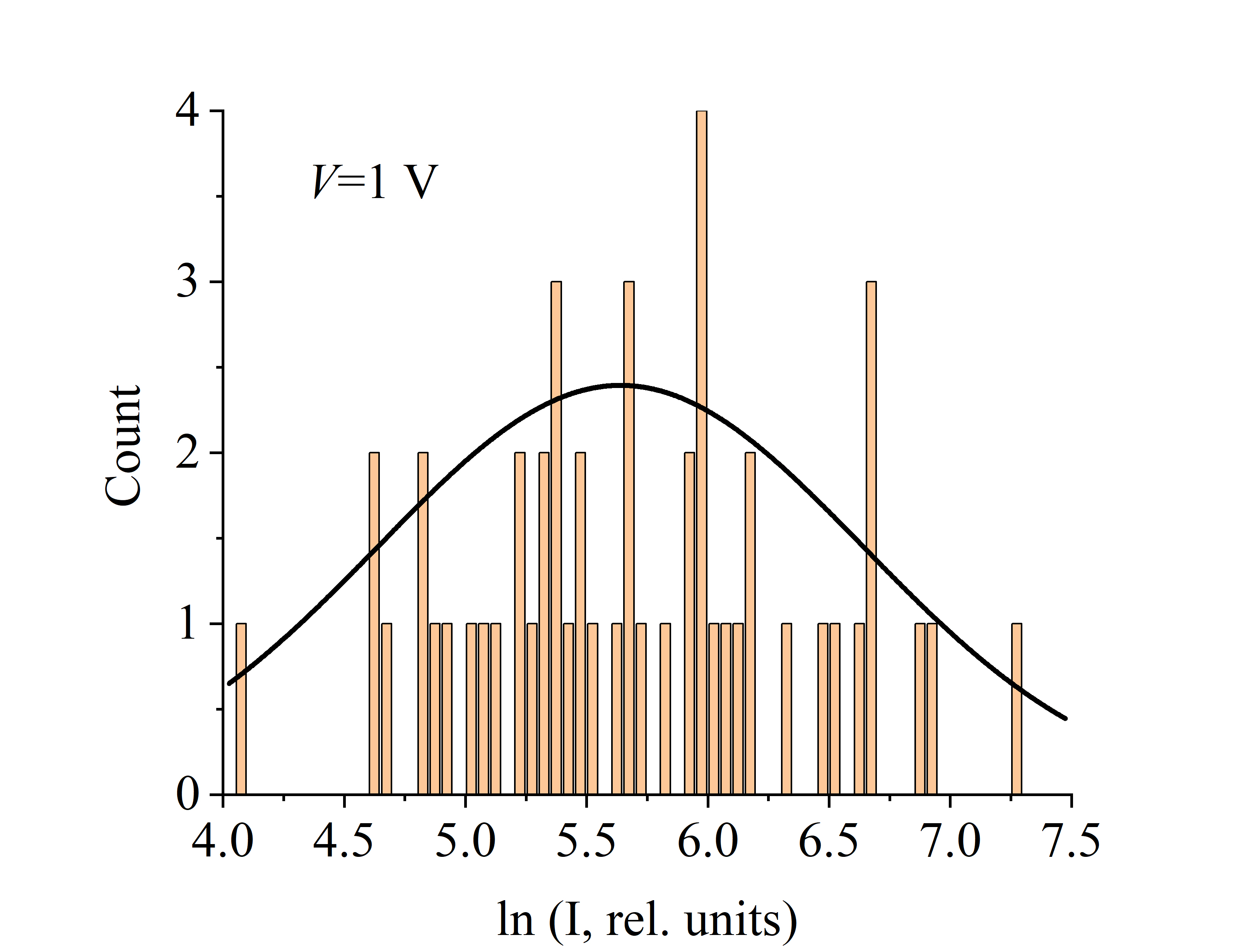}
\caption{The statistics of logarithms of currents for 50 random small (5x5) grids with $\xi _m=10$ and $kT=0.025$ eV under voltages $V=0.1$ V (left) and $V=1$ V (right). The lines represent fits by the normal distributions. \label{Fig:lognormals}}\end{figure}
\item As illustrated in Fig. \ref{Fig:map2}, the observed topology of conducting pathways in large grids with linear dimensions well above $L$ resembles random mesh and agrees with the standard images of percolation clusters. \cite{efros1986,stauffer1994}  The currents in nominally identical large grids with different disorder configurations are close to each other to the accuracy of $\sim 10$ \%, again, in agreement with the known results for percolation conduction.
\end{enumerate}

\begin{figure}[h!]
\includegraphics[width=0.47\textwidth]{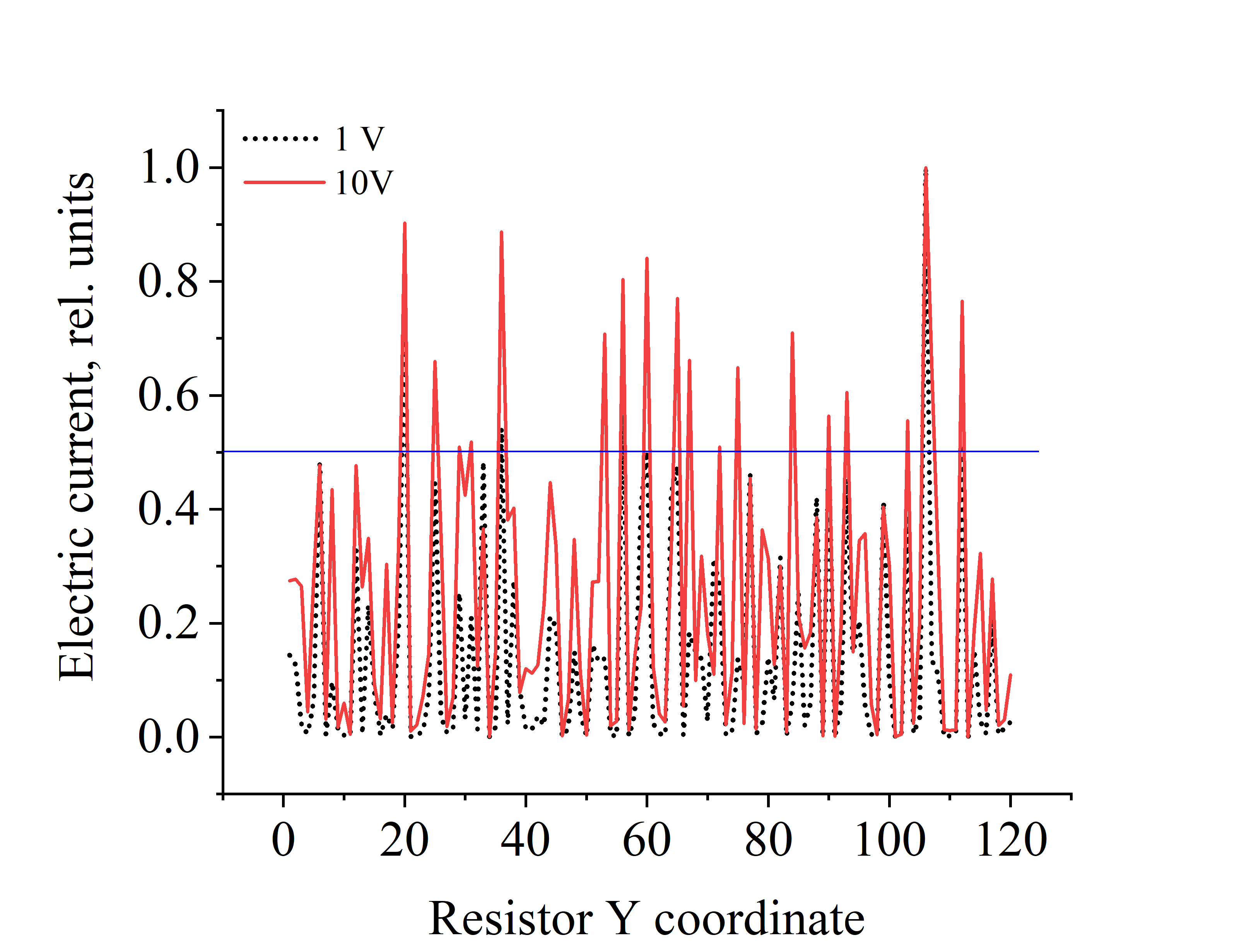}
\caption{Cross-sections of the current flow presented in Fig. \ref{Fig:map2} for the cases of $V=1$ V and $V=10$ V voltages normalized to their respective maximum values. The horizontal straight line represents our assigned level 0.5 of significance for peak counting as described in the text. \label{Fig:crosssec}}\end{figure}

\section{Discussion}\label{sec:disc}
We start our discussion with noting that the modeled very significant current increase (say, by $\sim 13$ orders of magnitude between 1 and 10 V voltages in Fig. \ref{Fig:lin_fit}) can hardly be observed experimentally. That discrepancy may be due to the fact that our modeling ignores Joule heat effects concentrating solely on the percolation aspects.

Considering the rectilinear type of conductive pathway topology in Fig. \ref{Fig:map1}, we note that it is consistent with the predictions of a more general analysis of transmittancy fluctuations in nonuniform barriers. \cite{raikh} The physics of it is that when the number of random resistors is relatively small, there is a significant probability that a few of them form a low resistance rectilinear path between the electrodes. Statistical variations between the resistances of such small systems are very significant and can exceed its average value as illustrated in Fig. \ref{Fig:lognormals}.

As the grid size increases, the probability of such paths exponentially decreases. Therefore, more complex winding paths forming a maze-like percolation cluster dominate the conduction. For sizes much greater than the correlation length, the system becomes effectively uniform and variations of resistances between nominally identical large systems become relatively small. \cite{efros}

Fig. \ref{Fig:crosssec} shows the distributions of horizontal (along the field lines) currents for two voltages arbitrarily taken at $X=40$ in Fig. \ref{Fig:map1}. These distributions are presented in the form normalized to their respective maximum values at $Y=108$ in Fig. \ref{Fig:crosssec} (before such a normalization, the current for 1 V is exponentially smaller than that of 10 V.) The two distributions turn out to be  statistically correlated, with correlation coefficient in the range of 80-85 \% for different cross-sections.
However, the number of significant peaks is noticeably larger for the $V=10$ V data. For example, arbitrarily taking the significance criterion as $I=I_{\max}/2$ for each of the two voltages yields the numbers of such peaks 17 for the 10 V data and 5 for the 1 V data. Therefore, the current cross-section graphs in Fig. \ref{Fig:crosssec} show, in agreement with the analytical predictions, \cite{shklovskii1979,patmiou2019} that the mesh size $L$ (evaluated from the average distance between the current spikes) decreases with voltage.

We note that the details of the current cross-sections and the numbers of significant peaks counted vary between different locations. In this study we did not collect a sufficient statistics to quantitatively confirm the analytical prediction that $L\propto 1/\sqrt{V}$.

Taken along with the above mentioned correlation between the two current cross-sections in Fig. \ref{Fig:crosssec}, our data imply that some of the conductive pathways that did not belong to the infinite cluster at lower voltages (1V), become its part as the voltage increases (to 10 V).

\section{Conclusions}\label{sec:concl}
We have developed a numerical model of non-ohmic percolation conduction explaining the Poole-Frenkel laws observed in a great variety of materials.
We would like to mention several extensions of our model.

While our algorithm assumes an infinite grid of resistors connecting two electrodes, it also sets a base for the placement of multiple electrodes, of interest in studying neuromorphic applications of percolation conduction.

Adding  the displacement currents (i. e. capacitive properties) our modeling will allow to study frequency dependent and pulse percolation, of interest for deploying the percolation conduction as a base for reservoir computing. \cite{karpov2020a}

Our modeling allows to introduce the algorithm of nonvolatile resistance changes in microscopic resistors of the grid (plasticity), which will extend its applicability over multi-valued memory applications.

Finally, taking into account the Joule heat generation and its feedback on the current voltage characteristics can have a significant effect on the macroscopic properties of percolation conduction, also possible with a proper extension of our modeling.

\newpage
\onecolumngrid
\appendix
\section*{Appendix}
PERCOLATION-SIMULATOR(N, $\xi$)

\begin{table}[hb]
\begin{tabular}{l l}
\textbf{in:\qquad\qquad\qquad}&N is the number of resistors in a row or column for an $N-by-N$ grid of resistors\\
&$(N \geq$ 4); $\xi$ is the largest possible number for randomly generated conductance\\
&values, such that $ \sim 10 \leq \xi \leq \sim 20$ and $\xi$ exists in the set of natural numbers\\
\textbf{out:\qquad\qquad}&returns the one-dimensional array \textit{I}, which contains the natural log of the total current\\
&for each value of the external volatage in the sequence $<1, 2, …, m>$\\

\textbf{constant: }&the maximum voltage to calculate current for $ m  = 10$; the middle node of\\
&grid for calculating the average current density, $n = \lceil N / 2 \rceil $; the convergence\\
&criterion for $\delta, e = 0.001 \cdot N $\\
\textbf{local:}&$G$ is the $N^2-by-N^2$ conductance matrix, setup as a two-dimensional array, where\\
&each element is base-e raised to the negative of a random number that exists in\\
&the set of rational numbers, between 0 - $\xi$; $F$ is the $N-by-N $ matrix that represents\\
&the exponentials of the nodal voltages, with voltage normalized in units of 2kT/e;\\
&$w$, $x$, and $y$ are the exponentials of the field components in the negative-x,\\
&positive-x, and y directions respectively, with the field normalized in units of\\
&2kT/(ea); $\delta$ is the relative current root-mean-square and must converge to a value\\
&less then $e$ before calculating $I_{\nu-1}$\\
\end{tabular}
\end{table}

\begin{spacing}{1.0}
 \enspace1.$|G| \leftarrow|N^2 \cdot N^2|$\hspace{2.5cm} //Reserve space for $G$, an $ N^2 x N^2$ matrix, as a 2D array.

 \enspace2.$G \leftarrow$ CALCULATE-CONDUCTANCE$(N^2, \xi)$

 \enspace3.$|I| \leftarrow |m|$\hspace{3.5cm}				//Reserve space to store current for voltages $<1, 2, …, m>$

\enspace4.\textbf{for }$\nu\leftarrow1...m$ \textbf{do}\hspace{2.4cm}//For each voltage in sequence $<1, 2,… m>$ calculate $I_{\nu-1}$

 \enspace5.\hspace{0.5cm}$w\leftarrow \exp((-1 \cdot \textnormal{v}) / (0.05 \cdot \textnormal{N}))$ //Calculate exponential of negative-x field component

\enspace6.\hspace{0.5cm}$x\leftarrow \exp(\textnormal{v}/(0.05 \cdot \textnormal{N}))$\hspace{1.3cm}//Calculate exponential of the positive-x field component

 \enspace7.\hspace{0.5cm}$y\leftarrow 1$ \hspace{3.6cm}//Calculate exponential of the y field component

 \enspace8.\hspace{0.5cm}$\delta\leftarrow 100$\hspace{3.4cm}//Set delta to an arbitrarily high value such that $\delta > e$

\enspace9.\hspace{0.5cm}$|F| \leftarrow |N \cdot N|$	\hspace{2.4cm}//Reserve space for $F$, an $NxN$ matrix, as a 2D array

 10.\hspace{0.5cm}\textbf{for } $i \leftarrow 0  … (N – 1)$ \textbf{do }\hspace{0.9cm}//Initialize first column of $F$

 11.\hspace{1cm}$F_{i,0} \leftarrow \exp(\textnormal{v}/ 0.05)$	\hspace{1.2cm}//$F_{i,0} =$ exp(eV/2kT) where 2KT/e at room temp, $\sim 0.05$

12.\hspace{0.5cm}\textbf{end for}

13.\hspace{0.5cm}\textbf{for} $i \leftarrow 0… (N – 1)$ \textbf{do}	\hspace{1cm}	//Initialize the rest of the elements in $F$ to exp(0)=1

 14.\hspace{1cm}\textbf{for} $j \leftarrow 0… (N – 1)$ \textbf{do}

15.\hspace{1.5cm}$F_{i,j} \leftarrow1$	\hspace{2.2cm}  //Last column will always retain value 1 because

16.\hspace{1cm}\textbf{end for }//potential will always be 0

17.\hspace{0.5cm}\textbf{end for }

18.\hspace{0.5cm}\textbf{while} $\delta>e$ \hspace{2.5cm} //Keep recalculating $F$ and $\delta$, until $\delta$ converges

19.\hspace{1cm}$F \leftarrow $ CALCULATE-F$(G, N, F, w, x, y)$

20.\hspace{1cm}$\delta\leftarrow$ CALCULATE-DELTA $(G, N, F, x, y)$

21.\hspace{0.5cm}\textbf{end while }

22.\hspace{0.5cm}// After calculating each nodal voltage in terms of the nodal voltage $(F)$, we now find the

 23.\hspace{0.5cm}// total current through a cross section of the grid, between column $ n$ and column $(n+1)$

 24.\hspace{0.5cm}\textbf{for }$ i \leftarrow 0… N-1$ \textbf{do }	\hspace{2.4cm}//Calculate current $I_{\nu-1}$

 25.\hspace{1cm}$a\leftarrow$ singleIndex$(i, n, (N,N))$	\hspace{1cm}// Get single index using $F$’s dimensions, $NxN$

26.\hspace{1cm}$b\leftarrow$ singleIndex$(i, n+1, (N,N))$
	
27.\hspace{1cm}$s \leftarrow (F_{i, n} \cdot  \exp(\textnormal{v} / (0.05 \cdot  N) ) ) / F_{i, n+1}$\hspace{1cm}//First term needed to calculate $I_{\nu-1}$

28.\hspace{1cm}$t \leftarrow (F_{i,n+1}/F_{i, n} \cdot \exp(\textnormal{v} / (0.05 \cdot  N) ) )$ \hspace{1cm}//Second  term needed to calculate $I_{\nu-1}$

29.\hspace{1cm}$I_{\nu-1}  \leftarrow I_{\nu-1} + G_{a, b} \cdot (s - t)$	

30.\hspace{0.5cm}\textbf{end for }

31.$\hspace{0.5cm}I_{\nu-1} \leftarrow \ln(I_{\nu-1} / 2)$

32.\textbf{end for }

33.return $I$
\end{spacing}

\twocolumngrid

\end{document}